\documentclass[10pt,aps,prb,twocolumn,amsmath,amssymb]{revtex4}
\usepackage{amsmath}
\usepackage{graphicx}
\usepackage{color}
\usepackage{multirow}

\newcommand{\maybedelete}[1]{{#1}}

\begin{document}

 \title{Combining internally contracted states and matrix product states to perform multireference perturbation theory}
\author{Sandeep Sharma}
\email{sanshar@gmail.com}
\affiliation{Max Planck Institute for Solid State Research, Heisenbergstra{\ss}e 1, 70569 Stuttgart, Germany\\
Department of Chemistry and Biochemistry, University of Colorado Boulder, Boulder, CO 80302, USA}
\author{Gerald Knizia}
\affiliation{Department of Chemistry, Pennsylvania State University, 401A Chemistry Bldg, University Park, PA 16802, USA}
\author{Sheng Guo}
\affiliation{Department of Chemistry, Princeton University, Princeton, NJ 08544, USA}
\author{Ali Alavi}
\email{a.alavi@fkf.mpg.de}
\affiliation{Max Planck Institute for Solid State Research, Heisenbergstra{\ss}e 1, 70569 Stuttgart, Germany\\
 Deptartment of Chemistry, University of Cambridge, Lensfield Road, Cambridge CB2 1EW, United Kingdom}
\begin{abstract}
We present two efficient and intruder-free methods for treating dynamic correlation on top of general multi-configuration reference wave functions---including such as obtained by the density matrix renormalization group (DMRG) with large active spaces.
The new methods are the second order variant of the recently proposed multi-reference linearized coupled cluster method (MRLCC) [S. Sharma, A. Alavi, J. Chem. Phys. 143, 102815 (2015)], and of N-electron valence perturbation theory (NEVPT2), with expected accuracies similar to MRCI+Q and (at least) CASPT2, respectively.
Great efficiency gains are realized by representing the first-order wave function with a combination of internal contraction (IC) and matrix product state perturbation theory (MPSPT).
With this combination, only third order reduced density matrices (RDMs) are required.
Thus, we obviate the need for calculating (or estimating) RDMs of fourth or higher order; these had so far posed a severe bottleneck for dynamic correlation treatments involving the large active spaces accessible to DMRG.
Using several benchmark systems, including first and second row containing small molecules, Cr$_2$, pentacene and oxo-Mn(Salen), we shown that active spaces containing at least 30 orbitals can be treated using this method. On a single node, MRLCC2 and NEVPT2 calculations can be performed with over 550 and 1100 virtual orbitals, respectively. 
We also critically examine the errors incurred due to the three sources of errors introduced in the present implementation - calculating second order instead of third order energy corrections, use of internal contraction and approximations made in the reference wavefunction due to DMRG. 
\end{abstract}
\maketitle

\section{Introduction}
Many processes in chemistry and biology involve catalysis by transition metal centers.
These include systems ranging from ubiquitous catalysts of organic synthesis to the active sites of most enzymes.
Unfortunately, the electronic structure of such systems is often \emph{strongly correlated}, which makes the theoretical study of their nature and reactivity difficult.
Strong correlation is characterized by the necessity of considering more than one Slater determinant to describe the electronic wavefunction \emph{even qualitatively}.
This property renders standard methods like Kohn-Sham Density Functional Theory (DFT) or single-reference wave function methods like MP2 or CCSD(T) unreliable, if not straight out inapplicable.
To describe such systems, multi-reference (MR) methods are needed.
However, currently available MR methods still have various defects, which frequently force researchers into uneasy compromises regarding active space size, quantitative accuracy, reliability, intruder states, or the need to strongly truncate the studied system to make computational expense manageable.
We here present new variants of multi-reference linearized coupled cluster (MRLCC2)\cite{Sharma2016,Sharma2014a,Sharma2015mrlcc} and $N$-electron valence perturbation theory (NEVPT2),\cite{Angeli2004,Angeli2006,Angeli2001,Angeli2002} which greatly alleviate these problems.
By judiciously combining internal contraction (IC)\cite{Meyer1977,Werner1990, Knowles1992,Celani2000,Celani2003} and matrix product state perturbation theory (MPSPT),\cite{Sharma2016,Sharma2014a,Sharma2015mrlcc} we arrive at MR methods capable of efficiently and accurately describing large and complex strongly correlated systems \emph{quantitatively} (i.e., including dynamic correlation, \emph{vide infra}) if combined with a suitable active space treatment, such as provided by DMRG reference functions.

To quantitatively describe a strongly correlated system, both dynamic correlation and static correlation must be accounted for.
However, both problems have rather different demands on the wave function ansatz.
To account for static correlation, a wave function ansatz needs high flexibility in many-body configuration space, but only low flexibility in one particle space---a small number of active orbitals is normally sufficient.
This problem is commonly treated with brute force full configuration interaction (FCI),\cite{Knowles1984} which is practical for up to approximately 16 active orbitals.
However, recently significant progress has been made with alternative methods to FCI; in particular, due to developments in the density matrix renormalization group (DMRG) algorithm, FCI Quantum Monte Carlo (FCIQMC),\cite{Cleland2010, Booth2009,Booth2013} and selected configuration interaction approaches\cite{Huron1973,Schriber2016,Holmes2016}, static correlation can now be treated accurately in much larger active spaces.
We here focus on DMRG,\cite{White1992,White1993} a variational method which minimizes the energy of a wavefunction parametrized as a matrix product state (MPS).\cite{Schollwock2011,chan2011}
DMRG can handle active spaces of around 30--40 orbitals, and in some cases even up to 100\cite{legeza-rev,C0CP01883J,kurashige,White1999,Chan2002,wouters14,Zgid2008,Nakatani2013,sharma:124121,Keller2015,Chan2016}.
However, by themselves the mentioned methods are not efficient for obtaining quantitative accuracy and for this dynamic correlation must also be calculated.

To account for dynamic correlation, a wave function ansatz capable of efficiently handling a \emph{large} set of (virtual) orbitals is required, while the effective many-body structure of the wave function can be comparatively simple and is amendable to convenient parameterization.
An efficient approach for this is to treat excitations out of a qualitatively correct, strongly correlated reference function.
Methods of this class are hierarchies based on truncated multireference configuration interaction (MRCI)\cite{wernercaspt2,Knowles1992,Shamasundar2011}, various flavors of perturbation theory\cite{caspt2,Fink2009,Angeli2001,Hirao1992} and coupled cluster theory.\cite{Lyakh2012,Evangelista2012} 
However, the established methods of these classes are based on CI treatments of the static correlation, which either makes them cumbersome to apply (RAS, GAS)\cite{Olsen1988,Malmqvist1990,Ma2011a}, or limits them to small active spaces (CAS).
Some alternative approaches fall outside this static-dynamic partitioning framework (e.g., methods based on geminals\cite{Surjan2012,Johnson2013}, Jastrow factors\cite{Neuscamman2012} or reduced density matrices\cite{mazziotti}); while promising, they have so far not matured into routinely applied methods.

For these reasons, a major challenge in recent years has been finding ways of combining the powerful DMRG approach to static correlation with one of the techniques more suitable for dynamic correlation (Sec.~\ref{sec:PreviousWork}).
It was only recently realized that second-order perturbation theories (PT2s) can be formulated with matrix product states (MPS) by recasting them into an effective optimization problem.\cite{Sharma2016,Sharma2014a,Sharma2015mrlcc}
This approach, called matrix product state perturbation theory (MPSPT), converges to the fully uncontracted version of a PT2, and can be performed with relatively minor modifications to an existing DMRG program.\cite{Sharma2016,Sharma2014a,Sharma2015mrlcc}
MPSPT can treat complex zeroth order Hamiltonians $\hat H_0$ involving two-body interactions; this allows formulating PT2 methods which are much more accurate than one-body-$\hat H_0$-based PT2s such as MP2 or CASPT2.
We recently introduced multi-reference linearized coupled cluster (MRLCC) as one such method.\cite{Sharma2015mrlcc,Sharma2016}
While formally a PT, the third order MRLCC3 was shown to be extremely accurate (comparable to MRCI+Q), stable with no intruder state problems, exactly size-consistent, and to a large extent able to compensate for sub-optimal reference states.\cite{Sharma2015mrlcc,Sharma2016}

However, the previous implementation of MRLCC is computationally still rather expensive; this is a result of parameterizing the perturbed state using an MPS, which cannot efficiently treat large virtual orbital spaces.
Here we solve this problem by parameterizing the perturbed wave function using a mixture of Celani-Werner-style partial internal contraction (IC)\cite{Celani2000,Celani2003} and MPSPT; and limiting ourselves to calculating only the second order correction to the energies.
The resulting method can effectively handle large virtual spaces, while at the same time being applicable to systems with 30 active orbitals and more. 
As a side product, we also obtain an efficient variant of NEVPT2, which can be applied to systems with 1000 virtual orbitals and 30 active orbitals at the same time without the need for invoking the strong-contraction approximation.

The article is organized as follows: Sec.~\ref{sec:PreviousWork} discusses MR methods of describing dynamic correlation, including their benefits and drawbacks, and previous work aiming to make them compatible with DMRG. Sec.~\ref{sec:ti} introduces the NEVPT2 and MRLCC2 theories, and covers our approach to their efficient implementation with IC and MPSPT.
Sec.~\ref{sec:vb} covers accuracy tests and benchmark applications;
in particular, applications to first row dimers, Cr$_2$, pentacene and oxo-Mn(Salen), are shown, and timing tests for MRLCC2 and NEVPT2 with different active spaces and basis set sizes are presented.
Sec.~\ref{sec:c} covers concluding remarks and outlook for future research.

\section{Parameterizing dynamic correlation: Problems and Solutions}\label{sec:PreviousWork}

The naive treatment of dynamical correlation in MR problems is extremely challenging,\cite{Buenker1974,Buenker1975} but a major breakthrough was made with the introduction of internal contraction (IC) by Meyer.\cite{Meyer1977} IC, now widely used in MR theories\cite{Werner1990,Mahapatra1998,Andersson1992,Evangelista2012,Angeli2002,dmrgct} aggressively truncates the many body basis and reduces the cost of the calculation from exponential to merely polynomial in the number of active space orbitals. The difficulty with IC is that the working equations become extremely cumbersome and can often only be derived using domain specific computer algebra systems. The working equations can be simplified by treating the active orbitals and the core orbitals on an equal footing as was done in the MRCI program developed by Werner-Knowles (WK scheme)\cite{Werner1990, Knowles1992}. But large efficiency gains can be achieved by recognizing that density matrices should only contain active space indices. This was done in the case of CASPT2 by Celani and Werner (CW scheme)\cite{Celani2000,Celani2003}, and much later for MRCI by Shamasundar et al\cite{Shamasundar2011}. 

The straightforward use of IC in perturbation theory or truncated configuration interaction with DMRG is complicated by the fact that fourth (and sometimes even higher) order reduced density matrices (RDM) in the active space are required. Calculation and storage of such high order RDMs is limited to an active space of around 25 orbitals beyond which it become prohibitively expensive\cite{kurashige_cr2,Guo2016,wouters:134110,Roemelt2016}. To circumvent this difficulty, some researches have resorted to approximating the higher body RDM by reconstructing its disconnected part by antisymmetric multiplication of lower body RDM and ignoring the density cumulant which is the connected part\cite{dmrgct,yanai:194106,Neuscamman2010,Zgid2009,dmrgmrci,Saitow2015}. The RDM reconstruction can be performed by setting the three body and four body cumulants to zero as was done for n-electron valence perturbation theory (NEVPT2)\cite{Zgid2009}; which resulted in severe numerical problems rendering the theory virtually unusable. Canonical Transformation (CT) theory\cite{dmrgct,yanai:194106,Neuscamman2010} also sets three and higher body cumulants to zero and is known to suffer from \emph{intruder} state problems (the CT intruder states are different than the ones in perturbation theory). A much milder approximation is to use the exact three body cumulants but setting the four body cumulants to zero\cite{dmrgmrci,Saitow2015}. This increases the numerical stability considerably but artifacts still remain that have to be eliminated by using level shifts. The underlying difficulty in neglecting cumulants beyond a certain rank is that, while the cumulants decay exponentially rapidly with the rank in weak correlation, they either decay very slowly or not at all in highly multireference situations\cite{Hanauer2012}.

In this work we follow a different route, by using \emph{partial} internal contraction. All terms that require 4-RDMs are treated fully uncontracted, and the rest of the terms, requiring three body or lower RDMs are treated using IC. This partitioning is in fact also used in the MRCI of Shamasundar et al.\cite{Shamasundar2011}. The reason this is a practical route is as follows: all terms which are difficult to treat using IC and require 4-RDMs (IC states with three active space indices and one virtual or core index) are relatively easily treated using uncontracted theories, because they involve only one core or virtual orbital, and are effectively singly excited states.
However, if these uncontracted terms are represented using wave functions, as done in previous work,\cite{Celani2000,Celani2003,Shamasundar2011} their number \emph{still} increases exponentially in the number of active orbitals.
For this reason, we here treat such terms with MPSPT,\cite{Sharma2016,Sharma2014a,Sharma2015mrlcc} which does not have this problem.
This combination of IC and MPSPT is the key development of the current work, which makes efficient MRLCC2 and NEVPT2 methods accessible to the large active spaces treatable with DMRG.
We will now discuss its theory and implementation.

\section{Theory and Implementation}\label{sec:ti}
\subsection*{Notation}
The molecular orbitals are divided into three subsets: the core orbitals indexed by $i, j, k, l$, which are always doubly occupied in the reference wavefunction $|\Psi_0\rangle$ (CASSCF or CAS-CI); the active orbitals $r,s,t, u$, which can have any occupancy in $|\Psi_0\rangle$; and the virtual orbitals $a,b,c, d$, which are unoccupied in $|\Psi_0\rangle$. General orbitals (i.e., the union of the three subsets) are indexed by $m,n,o,p$. We work exclusively with spin free quantities, by integrating out the spin degrees of freedom. Thus, the Hamiltonian is
\begin{align}
\hat{H} = t^n_m \hat{E}^m_n +  W^{mn}_{op} \hat{E}^{op}_{mn}.
\end{align}
Here and in the following, repeated indices imply contraction, operators are written with a hat (e.g. $\hat{E}$) and numerical tensors are written without a hat (e.g. $W$). $t^n_m$ are the one electron integrals,  $W^{mn}_{op} = \langle mn|op\rangle$ are the two electron integrals, and
\begin{align}
\hat{E}^m_n = \sum_{\sigma} \hat{a}_{m\sigma}^{\dag} \hat{a}_{n\sigma} && \hat{E}^{op}_{mn}=\sum_{\sigma\tau} \hat{a}_{o\sigma}^{\dag} \hat{a}_{p\tau}^{\dag} \hat{a}_{n\tau}\hat{a}_{m\tau}
\end{align}
are the spin-free single- and double excitation operators, in which $\sigma,\tau\in\{\alpha,\beta\}$ denote the spin degrees of freedom.

\subsection*{Definitions of $\hat{H}_0$}
In the Rayleigh-Schroedinger perturbation theory, the zeroth order wavefunction $|\Psi_0\rangle$ must be an eigenfunction of the zeroth order Hamiltonian $\hat H_0$. However, for a given $|\Psi_0\rangle$, this requirement does \emph{not} uniquely determine $\hat H_0$, and in general each viable choice of $\hat H_0$ will lead to a different perturbation theory. For example, the CASSCF (or CAS-CI) wavefunction is, at the same time, an eigenfunction of the Fock operator, Dyall's Hamiltonian,\cite{dyall} and Fink's Hamiltonian;\cite{Fink2009,Fink2006} choosing these as $\hat H_0$ leads to CASPT2 theory, NEVPT2 theory, and MRLCC2 theory, respectively, all of which have vastly different properties. In this work we focus on the Dyall- and Fink- Hamiltonians, which are described below. These have previously been shown to lead to accurate multi-reference perturbation methods which do not suffer from the intruder state problems plaguing CASPT2.\cite{Roos1995,Roos1996}
 
 Concretely, Dyall's Hamiltonian\cite{dyall} is given by
\begin{align}
\hat{H}_D &=  f_i^j \hat{E}^{i}_{j} + f^s_r \hat{E}^r_s + W^{rs}_{tu} \hat{E}^{tu}_{rs} + f_{a}^{b} \hat{E}^{a}_{b}, \label{eq:dyall}
\end{align}
where $f$ is a Fock matrix. $f$ contains modified one body integrals, approximating the interaction with core- and active electrons as a mean-field. It is defined as
\begin{align}
f_i^j &= t_i^j + 2W^{jk}_{ik} - W^{jk}_{ki} + \big(W^{jr}_{is} - W^{jr}_{si}\big) \Gamma^s_r \nonumber\\
f_a^b &= t_a^b + 2W^{ak}_{bk} - W^{ak}_{kb} + \big(W^{ar}_{bs} - W^{ar}_{sb}\big) \Gamma^s_r \nonumber\\
f_r^s &= t_r^s + 2W^{rk}_{sk} - W^{rk}_{sj},  \notag
\end{align}
where $\Gamma^r_s = \langle \Psi_0|\hat{E}^r_s|\Psi_0\rangle$ denotes the 1-RDM. 
When using CASSCF canonical orbitals, $f$ becomes diagonal, and the diagonal elements can be viewed as orbital energies. A diagonal Fock matrix (implicitly assumed in the work of Angeli et al.\cite{Angeli2004,Angeli2006,Angeli2001,Angeli2002}) simplifies NEVPT2 theory. However, in the present work we avoid this assumption, and the expressions derived here are more general.
 
 Fink's Hamiltonian\cite{Fink2009,Fink2006} is 
 \begin{align}
\hat{H}_F =\left. t^n_m \hat{E}^m_n +  W^{mn}_{op} \hat{E}^{op}_{mn}\right|_{\Delta n_{ex} = 0}, \label{eq:fink}
\end{align}
where $\Delta n_{ex} = 0$ signifies that all terms which would change the total number of electrons in the core, active or virtual spaces are removed.

An interesting property of Fink's Hamiltonian is that the zeroth order energy is equal to the energy of the reference wavefunction, and the first order energy is zero. This is not true in NEVPT2 theory; similarly to M{\o}ller-Plesset perturbation theory (MP2), in NEVPT2 the reference energy is given by the sum of the zeroth and first order energies, neither of which vanish. 

\subsection*{The first order wave function $|\Psi_1\rangle$}
 In perturbation theory, the first order correction ($\Psi_1$) to the reference wavefunction ($\Psi_0$) is determined by solving the linear equation 
 \begin{align}
 \big(\hat{H}_0 - E_0\big) |\Psi_1\rangle = -\hat{Q}\hat{V}|\Psi_0\rangle,\label{eq:solve}
 \end{align}
 in which $\hat V := \hat H - \hat H_0$ is the perturbation, and $\hat{Q}=1-|\Psi_0\rangle\langle\Psi_0|$ is a projector.
To define the representation of $|\Psi_1\rangle$, we first split the full $N$-electron Fock-space into configuration spaces characterized by a unique occupation pattern $(\Delta N_c,\Delta N_a,\Delta N_v)$. The three integers hereby denote the change in the \emph{total} number of electrons in core ($N_c$), active ($N_a$), and virtual ($N_v$) orbitals relative to $\vert\Psi_0\rangle$. With a two-body perturbation $\hat V$, eight such spaces can be reached by $\hat V\vert\Psi_0\rangle$, and are therefore required in the expansion of $|\Psi_1\rangle$ (Eq.~\eqref{eq:solve}). These spaces are listed in Table~\ref{tab:eight}, and their representation will be discussed next.
Neither Dyall's nor Fink's Hamiltonian ($\hat{H}_0$) change the number of electrons in the three orbital spaces; consequently, both are block-diagonal with respect to the eight configuration spaces. 
Therefore Eq.~\eqref{eq:solve} can be solved in each space independently.
\begin{table}

\caption{The eight classes of perturber states \maybedelete{reachable by applying a two-body perturbation $\hat V$ to the zeroth order state $|\Psi_0\rangle$, and} which therefore contribute to $|\Psi_1\rangle$. The classes are formed by changing the total number of electrons in core, active, and virtual orbitals, relative to the occupation pattern of $|\Psi_0\rangle$ (indicated by positive/negative numbers). Column 5 and 6 denote the basis of the many-body states and the order of the active-space RDM needed, \emph{if} the space were parameterized using internal contraction (IC). MPSPT denotes matrix product state perturbation theory. In the methods discussed here, the contribuitions to $|\Psi_1\rangle$ within spaces I--VIII can be solved for independently of each other.}\label{tab:eight}  
\smallskip
\begin{tabular}{lccccc@{\hspace*{2ex}}c}
\hline\hline
&\multicolumn{3}{c}{Occupation-change}&Basis in&Needed&Solution\\\cline{2-4}
&Core&Active&Virtual&IC case&RDM&strategy\\
\hline
I&$-$2&$~~$0&+2&$\hat{E}^{a}_{i}\hat{E}^{b}_{j}\vert\Psi_0\rangle$&1&IC\\
II&$-$1&$-$1&+2&$\hat{E}^{a}_{i}\hat{E}^{b}_{r}\vert\Psi_0\rangle$&2&IC\\
III&$-$2&+1&+1&$\hat{E}^{a}_{i}\hat{E}^{r}_{j}\vert\Psi_0\rangle$&2&IC\\
IV&$~~$0&$-$2&+2&$\hat{E}^{a}_{r}\hat{E}^{b}_{s}\vert\Psi_0\rangle$&3&IC\\
V&$-$2&+2&$~~$0&$\hat{E}^{r}_{i}\hat{E}^{s}_{j}\vert\Psi_0\rangle$&3&IC\\
VI&$-$1&$~~$0&+1&$\{\hat{E}^{a}_{i}\hat{E}^{r}_{s}\vert\Psi_0\rangle$, &3&IC\\
  &  & & &$\hat{E}^{a}_{s}\hat{E}^{r}_{i}\vert\Psi_0\rangle$, \\
  &  & & &$\hat{E}^{a}_{i}\vert\Psi_0\rangle\}$\\
VII&$-$1&+1&$~~$0&$\hat{E}^{r}_{t}\hat{E}^{s}_{j}\vert\Psi_0\rangle$&4$^*$&MPSPT\\
VIII&$~~$0&$-$1&+1&$\hat{E}^{r}_{s}\hat{E}^{a}_{t}\vert\Psi_0\rangle$&4$^*$&MPSPT\\
\hline\hline
\end{tabular}
\begin{flushleft}\small{$^*$In this work no 4-RDM is required because we do \emph{not} use internal contraction for classes VII and VIII.}\end{flushleft}
\end{table}
 
To solve Eq.~\eqref{eq:solve}, we first choose a suitable basis of many-body states $\{\Phi_{\mu}\}$, covering the eight configuration classes. We then write the first order wave function as a linear combination
\[ |\Psi_1\rangle = d^{\nu}|\Phi_{\mu}\rangle. \]
The choice of the basis states $\{\Phi_{\mu}\}$ is not unqiue; various approaches are possible.
For example, one could choose to include \emph{all} individual determinants (or CSFs) which have non-zero overlap with $\Psi_1$; this is known as an uncontracted formulation of the theory. Unfortunately, with this choice the cost of solving Eq.~\eqref{eq:solve} generally scales exponentially with the number of active orbitals, quickly becoming impractical.
Alternatively, the excitation classes can be represented using internal contraction (IC). Here the many-body basis $\{\vert\Phi_{\mu}\rangle\}$ is formed by acting with orbital excitation operators $\hat{O}_{\mu}$ (e.g. $\hat{O}_{\mu} = \hat{E}^{a}_{i}\hat{E}^{b}_{j}$) onto the \emph{entire} zeroth order state $|\Psi_0\rangle$, rather than its individual configurations:
\begin{align}
   {\vert\Phi_{\mu}}\rangle = \hat{O}_{\mu}\vert\Psi_0\rangle.\label{eq:icbasis1}
\end{align}
A list of such states suitable for representing $|\Psi_1\rangle$ is shown in column 6 of Table~\ref{tab:eight}. With IC, the parameterization of $|\Psi_1\rangle$ becomes much more efficient than in uncontracted theories; the cost of solving Eq.~\eqref{eq:solve} reduces from exponential to polynomial in the number of active orbitals. The drawback of this approach is that the IC basis states \eqref{eq:icbasis1} are complicated, non-orthogonal many-body states. This leads to complex expressions for operator matrix elements between IC basis states; these generally contain series of tensor contractions involving high order RDMs, as shown in the next section.


\subsection*{Internal contraction}
To demonstrate the positive and negative aspects of IC, we here derive the equations for perturber space IV (see Table~\ref{tab:eight}). We write both $\Psi_1$'s and $\hat{V}|\Psi_0\rangle$'s space IV contribution as linear combination of IC states,
\begin{align}
   \vert\Psi_1^{IV}\rangle &=  d^{rs}_{ab} \hat{E}^{a}_{r}\hat{E}^{b}_{s}|\Psi_0\rangle 
\\ (\hat V \vert\Psi_0\rangle)^{IV} &= W^{tu}_{cd}\hat{E}^{c}_{t}\hat{E}^{d}_{u}|\Psi_0\rangle.
\end{align}
Here $d^{sr}_{ab}$ are the unknown parameters of $\vert{\Psi_1}\rangle_{IV}$, and $W^{tu}_{cd}$ are the two-body integrals.
Substituting into Eq.~\ref{eq:solve}, we get
\begin{align}
 \big(\hat{H}_0 - E_0\big) d^{rs}_{ab} \hat{E}^{a}_{r}\hat{E}^{b}_{s}|\Psi_0\rangle = -Q W^{tu}_{cd}\hat{E}^{c}_{t}\hat{E}^{d}_{u}|\Psi_0\rangle.
\end{align}
To obtain closed numerical equations for $d^{sr}_{ab}$, we left-project onto the bra states $\langle\Psi_0|\hat{E}^{r'}_{a'}\hat{E}^{s'}_{b'}$, arriving at 
\begin{align}
\langle\Psi_0|\hat{E}^{r'}_{a'}\hat{E}^{s'}_{b'} (\hat{H}_0 - E_0)& \hat{E}^{a}_{r}\hat{E}^{b}_{s}|\Psi_0\rangle d^{rs}_{ab} =\nonumber\\
 &-\langle\Psi_0|\hat{E}^{r'}_{a'}\hat{E}^{s'}_{b'}\hat{E}^{c}_{t}\hat{E}^{d}_{u}|\Psi_0\rangle W^{tu}_{cd}\nonumber.
\end{align}
With $A$ and $S$ as defined in Eqs.~\eqref{eq:Awithlotsofindices} and \eqref{eq:Swithlotsofindices}, this can be equivalently written in tensor form as
\begin{align}
A^{r'a's'b'}_{rasb} d^{rasb} =& -S^{r'a's'b'}_{tcud} w^{tcud} \label{eq:contract}.
\end{align}

The tensor $S$ represents an overlap matrix of IC states
\begin{align}
   S^{r'a's'b'}_{tcud} = \langle\Psi_0|\hat{E}^{r'}_{a'}\hat{E}^{s'}_{b'}\hat{E}^{c}_{t}\hat{E}^{d}_{u}|\Psi_0\rangle.\label{eq:Swithlotsofindices}
\end{align}
Using Wick's theorem, this expression can be simplified into a sum of products of one 2-RDM with only active indices $rstu$, and two Kronecker-$\delta$s connecting the virtual indices $abcd$ (see Eq.~\eqref{eq:overlap}). Similarly, $A$ (Eq.~\eqref{eq:Awithlotsofindices}) can be expressed as a sum of tensor contractions via Wick's theorem; these contain interaction tensors ($W$ and $t$), $\delta$-symbols with virtual and/or core indices, and up to four-body RDMs (with up to eight active indices---four from $\hat H_0$ and four from $\vert\Psi_1^{IV}\rangle$). However, the 4-RDM can be eliminated, \emph{without approximation}, as follows:
\begin{align}
A^{r'a's'b'}_{rasb} &= \langle \hat{E}^{r'}_{a'}\hat{E}^{s'}_{b'} (\hat{H}_0 - E_0) \hat{E}^{a}_{r}\hat{E}^{b}_{s}\rangle\label{eq:Awithlotsofindices}
\\ &= \langle \hat{E}^{r'}_{a'}\hat{E}^{s'}_{b'} \big[(\hat{H}_0 - E_0), \hat{E}^{a}_{r}\hat{E}^{b}_{s}\big]\rangle \nonumber
\\ &\qquad+\langle \hat{E}^{r'}_{a'}\hat{E}^{s'}_{b'} \hat{E}^{a}_{r}\hat{E}^{b}_{s}(\hat{H}_0 - E_0) \rangle \label{eq:comm1}
\\ &= \langle \hat{E}^{r'}_{a'}\hat{E}^{s'}_{b'} \big[(\hat{H}_0 - E_0), \hat{E}^{a}_{r}\hat{E}^{b}_{s}\big]\rangle \label{eq:comm2}
\end{align}
To cancel the second term of \eqref{eq:comm1}, we used that $|\Psi_0\rangle$ is an eigenfunction of $\hat{H}_0$ with eigenvalue $E_0$.
For evaluating the commutator in Eq~\eqref{eq:comm2}, only up to three-body RDMs are needed. Using similar procedures, IC equations for the other classes in Table~\ref{tab:eight} can be derived; the maximum order RDM required in these equations is shown in column 6. 
Since computing and storing a 4-RDM would be costly, we treat only the indicated six of the eight perturbation classes with IC.
\\
\\
\noindent\emph{Wick's theorem and tensor contractions}\\
To arrive at practical working equations for the $\Psi_1$ contributions in the various perturbation classes, the second-quantized operator expressions, such as
\[ \langle \hat{E}^{r'}_{a'}\hat{E}^{s'}_{b'} \big[(H_0 - E_0), \hat{E}^{a}_{r}\hat{E}^{b}_{s}\big]\rangle\]
of Eq.~\eqref{eq:comm2}, must be transformed into a series of tensor contractions using Wicks' theorem.
While this processes is relatively straight-forward, it is extremely tedious and error-prone when done manually. For this reason, various computer programs have been developed to automate the evaluation process. The resulting expressions only contain one- and two-body integral tensors ($t$ and $W$), $\delta$-symbols, and RDMs containing exclusively active indices. In this work we used the freely available program SecondQuanizationAlgebra, which was originally developed for implementing Canonical Transformation theory,\cite{Neuscamman2009} and has since been modified and extended.\cite{dmrgmrci} For example, the program evaluates the left hand side of Eq.~\ref{eq:contract}, where $A$ is given by \eqref{eq:comm2} with $\hat{H}_0$ being Fink's Hamiltonian, into a sum of eleven different tensor contraction terms. The first three of these are 
\begin{align}
c^{r'a's'b'} &= A^{r'a's'b'}_{rasb} d^{rasb}\nonumber\\
 &= -4t^{t}_{r} \Gamma^{r's'}_{ts} d^{rasb} \delta_a^{a'} \delta_{b}^{b'} \nonumber\\
&\qquad+ 4t^{a'}_{b} \Gamma^{r's'}_{sr} d^{rasb} \delta_{a}^{b'} \nonumber \\
&\qquad+ 4W^{ta'}_{t'b}   d^{rasb}\Gamma^{r's't'}_{srt} \delta_{a}^{b'}+ \cdots, \label{eq:aavv}
\end{align}
The full equations for all six perturbation classes treated with IC, for both the Fink's and Dyall's zeroth order Hamiltonians, are given in the supplementary material.

Once the equations have been derived, they have to be factorized into a sequence of binary tensor contractions (contractions in which one output tensor is formed by contracting the common indices of \emph{two} input tensors). At this moment our program does not attempt to find the optimal factorization for a set of tensor contractions; while a powerful factorization algorithm has been reported in the literature,\cite{Engels-Putzka2011} it is so far not applicable to our case, and not easy to implement. Instead we factorize each multi-tensor-contraction independently using a simple heuristic.\footnote{For a contraction with $n$ input tensors ($n>2$), our heuristic algorithm first choses the two input tensors which, if contracted over their common indices, produce the smallest intermediate tensor. This binary contraction is then evaluated, and the two input tensors are replaced by the single output tensor, resulting in a $n-1$ tensor contraction. This is repeated until a binary contraction is reached, which is evaluated with transpositions and matrix multiplications.} Several computer libraries capable of efficiently evaluating the final sequence of binary contractions have recently become available.\cite{Epifanovsky2013,6569864,Calvin2015,doi:10.1021/jp034596z} We here use a prototyping library developed by one us, which will be described elsewhere.\\
\\
\noindent\emph{Solution of the linear equations}\\
In general, the IC basis functions \eqref{eq:icbasis1} of a given excitation class are not orthogonal to each other.
For example, for perturber class IV, the overlap matrix is 
\begin{align}
S^{r'a's'b'}_{tcud} &= \langle \hat{E}^{r'}_{a'}\hat{E}^{s'}_{b'}\hat{E}^{c}_{t}\hat{E}^{d}_{u} \rangle \notag\\
&=  \Gamma^{r's'}_{tu} \delta^{a'}_{c} \delta^{b'}_{d} + \Gamma^{s'r'}_{tu} \delta^{b'}_{c} \delta^{a'}_{d}. \label{eq:overlap}
\end{align}
An orthogonal basis can be generated by limiting the virtual indices in $\hat{E}^{c}_{t}\hat{E}^{d}_{u}|\Psi_0\rangle$ to $c\geq d$, and diagonalizing the two body density matrix $\Gamma^{r's'}_{tu}$ (just the first term in Eq~\eqref{eq:overlap}). With this orthogonal basis, the metric $S$ can be eliminated from Eq.~\eqref{eq:contract}, simplifying it to 
\begin{align}
\tilde{A}^{\alpha' a'b'}_{\alpha ab} \tilde{d}^{\alpha ab} =&  \tilde{w}^{\alpha' a'b'}, \label{eq:ortho}
\end{align}
where we have defined the orthogonal-basis quantities
\begin{align}
\tilde{A}^{\alpha' a'b'}_{\alpha ab} =& U^{\alpha'r's'}A^{r'a's'b'}_{rasb}U^{\alpha rs} \nonumber\\
\tilde{d}^{\alpha ab} =& U^{\alpha rs}d^{rasb}\nonumber\\
\tilde{w}^{\alpha ab} =& U^{\alpha rs}w^{rasb}. 
\end{align}
Here $U^{\alpha rs}$ is the $\alpha^{th}$ eigenvector of the 2-RDM. We use the conjugate gradient (CG) method for solving Eq.~\eqref{eq:ortho}. In each CG iteration, $\tilde{A}^{\alpha' a'b'}_{\alpha ab} \tilde{d}^{\alpha ab} $ must be evaluated, which is non-trivial. Using Wick's theorem, we only arrive at tensor contractions for $A^{r'a's'b'}_{rasb} d^{rasb}$ in the \emph{non}-orthogonal basis (note the missing $^\sim$s). Thus, we first convert $\tilde{d}$ to $d$ by left multiplying by $U^\dagger$. Once $d$ is calculated, $c = Ad$ is evaluated in the non-orthogonal IC basis as described in the previous section. Finally, $c$ is converted to $\tilde{c}$ by left multiplying by $U$. Thus, the entire CG scheme is performed in the orthogonal basis, and only the contraction $c = Ad$ is evaluated in the non-orthogonal IC basis.

A simpler solution scheme is possible if using Dyall's Hamiltonian and canonical orbitals. In this case, the product $c^{r'a's'b'} =A^{r'a's'b'}_{rasb} d^{rasb}$
can be simplified to $c^{r'as'b} = (K^{r's'}_{rs}+f_a^a+f_b^b)d^{rasb}$. Note that the virtual orbital indices are identical in tensors $c$ and $d$. This simplification allows a closed-form solution of the linear equation, making an iterative CG scheme unnecessary. For this, only the $K$ matrix must be diagonalized, which is of the same size as the overlap matrix $S$.

\subsection*{Matrix product state perturbation theory}
As noted in Table~\ref{tab:eight}, if we were to employ IC for perturber classes VII and VIII, this would require computing and storing the 4-RDM of the reference function $|\Psi_0\rangle$. This would be costly, and would severely limit the applicability of our methods. To avoid this problem, we represent these perturber classes with matrix product state perturbation theory (MPSPT) instead of IC. The MPSPT approach requires no RDMs, and is capable of representing arbitrary \emph{uncontracted} wave functions of the respective perturber classes. The core idea of this approach is to write the linear equation~\eqref{eq:solve} as a minimization problem using the Hylleraas functional:
\begin{align}
H[\Psi_1] = \langle\Psi_1|H_0 - E_0 |\Psi_1\rangle + 2\langle\Psi_1|QV|\Psi_0\rangle.
\end{align}
This reformulation makes Eq.~\eqref{eq:solve} susceptible to the optimization techniques of matrix product state (MPS) ansatz wave functions.

\begin{figure}
\begin{center}
\includegraphics[width=0.4\textwidth]{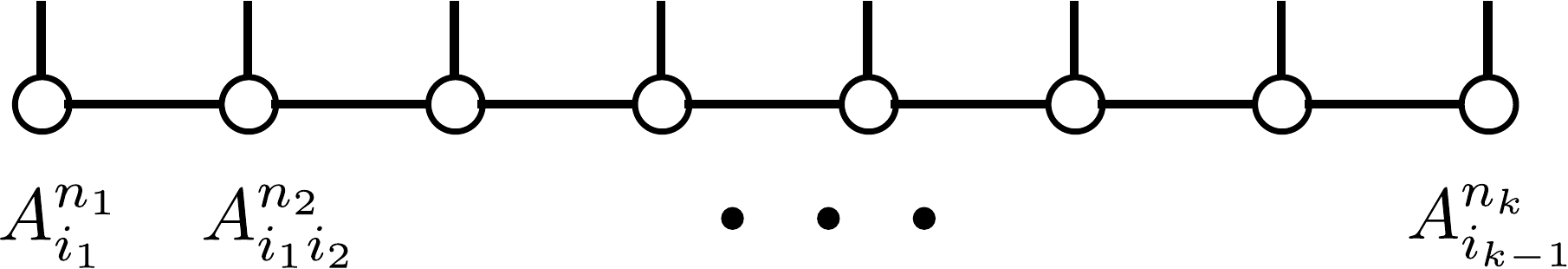}
\end{center}
\caption{A matrix product state (MPS) can be represented graphically using a series of three-dimensional tensors $A^{n_l}_{i_{l-1} i_l}$. In these,
the physical index $n_l$ (pointing upwards) denotes the occupation of the orbital $l$, and the other two indices $i_l, i_{l-1}$ (pointing horizontally, and known as virtual indices), are sequentially contracted.}\label{fig:mps}
\end{figure}

In this work both the reference wavefunction $|\Psi_0\rangle$ and the first order correction $|\Psi_1\rangle$ are parametrized using a MPS ansatz. In an MPS ansatz, the wavefunction is expressed as a product of a set of three-dimensional tensors, which is shown graphically in Fig.~\ref{fig:mps}. The virtual bonds (horizontal lines in Fig.~\ref{fig:mps}) encode entanglement between the different orbitals, and by increasing this bond dimension ($M$), any wavefunction can be exactly represented using an MPS. An important property of the MPS ansatz is that overlaps and operator expectation values between two different MPS wave functions can be calculated at a polynomial cost in the number of orbitals ($k$) and the virtual bond dimension ($M$). A summary of the costs of each such calculation is shown in Table~\ref{tab:mps}.

\begin{table}
\caption{Scaling of the computational cost for calculating the overlap and expectation values of one body ($\hat O_1$) and two body ($\hat O_2$) operators between two MPS $|\Phi_1\rangle$ and $|\Phi_2\rangle$. Here $k$ is the number of orbitals, and $M_1$ and $M_2$ are the virtual bond dimensions of $|\Phi_1\rangle$ and $|\Phi_2\rangle$ respectively.}\label{tab:mps}
\smallskip
\begin{tabular}{l@{\hspace*{1em}}l}
\hline
\hline
Operation& Computational Cost\\
\hline
$\langle\Phi_1|\Phi_2\rangle$ & $kM_1^2M_2 + kM_1M_2^2$\\
$\langle\Phi_1|\hat{O}_1|\Phi_2\rangle$ & $k^2M_1^2M_2 + k^2M_1M_2^2$\\
$\langle\Phi_1|\hat{O}_2|\Phi_2\rangle$ & $k^3M_1^2M_2 + k^3M_1M_2^2$\\
\hline\hline
\end{tabular}
\end{table}

\begin{figure}
\begin{center}
\includegraphics[width=0.5\textwidth]{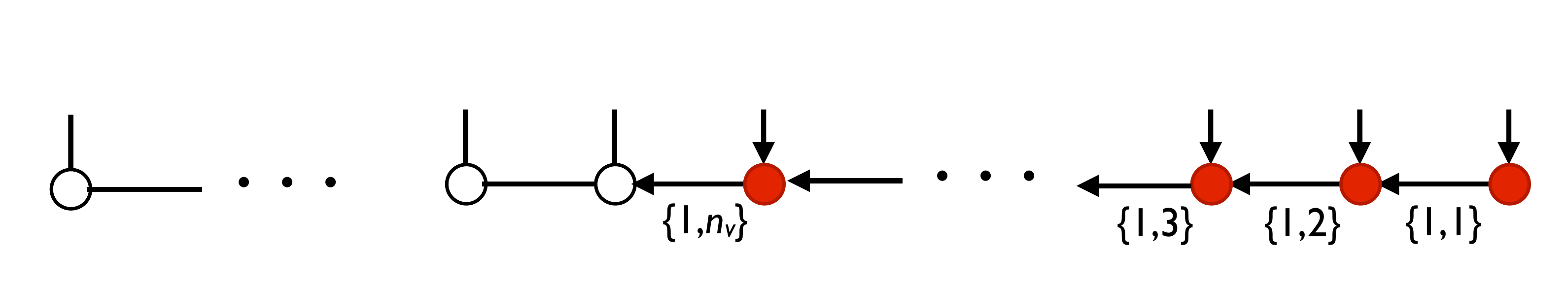}
\end{center}
\caption{The orbitals in the red are the virtual orbitals and are ordered such that they are all together to the right of the active space orbitals shown with empty circles. These virtual orbitals contain a maximum of one electron for the first order state corresponding to perturber class VIII in table~\ref{tab:eight}. Given this constraint, the virtual bond dimension of the $n^{th}$ tensor starting from the right most virtual orbital has a left bond dimension of $n+1$, with one state representing empty virtual orbitals and $n$ single electron states. In the above figure the single zero electron and $n$ single electron states are represented as \{$1,n$\}.}\label{fig:bigdot}
\end{figure}

The ability to perform the various linear algebra operations shown in Table~\ref{tab:mps} in polynomial time makes it possible to calculate the Hylleraas functional, and to subsequently optimize it by varying the tensors parameterizing the MPS $|\Psi_1\rangle$. The optimization is performed using a sweep algorithm, just as DMRG optimizations of ground state wavefunctions. In a sweep iteration, each single tensor $A^{n_l}_{i_{l-1} i_l}$ of $|\Psi_1\rangle$ is optimized individually, while keeping other $A$ tensors fixed. This is repeated until convergence of all $A$s. So this sweep algorithm solves the optimization problem of a large, multi-linear function of $k$ tensors by iteratively solving a single linear equation for a single tensor\cite{Sharma2016,Sharma2014a,Sharma2015mrlcc} The simple MPSPT algorithm requires a sweep over the entire set of orbitals and optimizes each tensor separately. This can become prohibitively expensive when working with several hundred virtual orbitals. But by realizing that the first order state has a maximum of a single electron in the virtual space, one can once and for all fix the tensors corresponding to the virtual orbitals, and subsequently no sweeps over these orbitals are necessary. These virtual orbital tensors are shown in red in the figure~\ref{fig:bigdot}; they are fixed at the beginning of the calculation and are not changed for the rest of the calculation. Note that the maximum virtual bond dimension required for these tensors is $n_v+1$, where $n_v$ is the number of virtual orbitals, which is less than a typical $M$ of a few thousand that is routinely used in calculations.

A similar simplification is possible for the perturber class VII, where instead of a single electron in the virtual orbitals, the core orbitals have a single hole. Thus the tensors corresponding to the core orbitals can be fixed once at the beginning of the calculation and require a maximum bond dimension of $n_c+1$, where $n_c$ is the number of core orbitals. 

In the case of NEVPT2 and canonical virtual orbitals, the first order correction to the wavefunction due to perturber class VIII can be written as $\sum_a |\Psi_1^a\rangle$, where in the state $|\Psi_1^a\rangle$ the virtual orbital $a$ is singly occupied; and each of these $n_v$ states can be calculated independently. Similarly, the perturber class VII can be split into $n_c$ states that can be determined independently when canonical core orbitals are used. Thus, using canonical orbitals in a NEVPT2 offers significant practical advantages, and allows working with very large core and virtual orbital spaces possible.

\section{Validation and Benchmarks}\label{sec:vb}

\begin{table*}
\caption{Atomization energies (AE), ionization potential (IP) and electron affinities (EA) of various benchmark molecules are calculated using uncontracted MRLCC3,  uncontracted MRLCC2 and internally contracted MRLCC2 implemented in the present work. Where possible we have also tabulated accurate estimates of the FCI energies; AE were calculated using FCIQMC\cite{Cleland2012}, whereas IP and EA were calculated using DMRG. The FCIQMC energies and geometries of molecules used for AE calculations are obtained from work by Cleland et al.\cite{Cleland2012} and the geometries for molecules used for calculating IP and EA were obtained from work by Knizia et al.\cite{knizia:054104}.}\label{tab:iclcc}
\begin{tabular}{lccccccccccccccccc}
\hline
\hline
&&\multicolumn{3}{c}{FCI}&&\multicolumn{3}{c}{UMRLCC3}&&\multicolumn{3}{c}{UMRLCC2}&&\multicolumn{3}{c}{p-ICMRLCC2}\\
\cline{3-5}\cline{7-9}\cline{11-13}\cline{15-17}
Molecule&&~~~DZ~~~&~~~TZ~~~&~~~QZ~~~&&~~~DZ~~~&~~~TZ~~~&~~~QZ~~~&&~~~DZ~~~&~~~TZ~~~&~~~QZ~~~&&~~~DZ~~~&~~~TZ~~~&~~~QZ~~~\\
\cline{1-1}\cline{3-5}\cline{7-9}\cline{11-13}\cline{15-17}
\multicolumn{7}{l}{\emph{Atomization Energy}}\\
C$_2$&&-0.207&	-0.223&	-0.228&&-0.204	&-0.220	&-0.226&&-0.204&	-0.216&	-0.221	&&	-0.204&	-0.216&	-0.221\\
N$_2$&&-0.320&	-0.346&	-0.356&&-0.317	&-0.343	&-0.354&&-0.314	&-0.337	&-0.347	&&	-0.314	&-0.336	&-0.346\\
F$_2$&&-0.044&	-0.057&	-0.059&&-0.040	&-0.052	&-0.056&&-0.051	&-0.064	&-0.067	&&	-0.051	&-0.063	&-0.066\\
\\\multicolumn{7}{l}{\emph{Ionization Potential}}\\
H$_2$O&&-0.674 &- &- &&-0.672&-0.684&-0.689&&-0.674&-0.684&-0.689&&0.674&-0.684&-0.689\\
NH$_3$&&-0.618 &- & -&&-0.617&-0.624&-0.628&&-0.612&-0.617&-0.620&&	-0.612&-0.617&	-0.620\\
Cl$_2$&& -&- &- &&-0.412&-0.415&-0.420&&-0.411&-0.412&-0.416&&-0.411&-0.413&-0.416\\
OH&&-0.449 &- & -&&-0.448&-0.467&-0.474	&&	-0.470&	-0.452&-0.477	&&-0.452&-0.470&		-0.470\\
\\\multicolumn{8}{l}{\emph{Electron Affinity}}\\
CH$_3$&&-0.048 &- &- && -0.051&-0.031&	-0.021&&	-0.049	&-0.030	&-0.021&&	-0.049	&-0.031&	-0.022\\
CN&&0.100 &- &- &&0.099&0.124&0.132&&0.100&0.124&0.133&&0.100&0.124&0.133\\
NO&&-0.053 &- &- &&-0.054&-0.029&-0.017&&-0.048&-0.022&-0.009&&-0.048&-0.022&-0.010\\
SH&& -&- &- &&0.040&0.064&0.075&&0.039&0.061&0.071&&0.039&0.061&0.071\\
\hline
\hline
\end{tabular}
\end{table*}

\begin{table}\label{tab:lcc}
\caption{Effect of internal contraction and accuracy of $\vert \Psi_0\rangle$ on absolute MRLCC2 energies.
The rows show reference energy ($E_0+2099.0$), and MRLCC2 correction due to classes I-VI of Tab.~\ref{tab:eight}($E_{2,\mathrm{IC}}$), and classes VII-VIII ($E_{2,\mathrm{MPTSPT}}$), in atomic units.
The UC (uncontracted) column contains reference energies obtained with uncontracted basis states for all classes. In the other four columns, an IC basis is used for perturber classes I-VI, and classes VII-VIII are represented with MPSPT. In these four columns, $\vert \Psi_0\rangle$ is an MPS with maximum virtual bond dimension $M$ of 500, 80, 50 and 30 respectively. This illustrates the effect of approximating $\vert \Psi_0\rangle$ on the correlation energy.}
\smallskip
\begin{tabular}{lcccccc}
\hline\hline
&UC&~&\multicolumn{4}{c}{Bond dimension $M$ of $\vert \Psi_0\rangle$}\\\cline{4-7}
&&&500& 80 & 50 & 30 \\
\cline{1-2}
\cline{4-7}
$E_0$ & -0.2499&&-0.2499&-0.2487&-0.2461&-0.2192\\
$E_2$\tiny{(I--VI)}$^a$& -1.0597&&-1.0578&-1.0582&-1.0587&-1.0629\\
$E_2$\tiny{(VII--VIII)}$^b$& -0.0672&&-0.0672&-0.0675&-0.0679&-0.0709\\
\hline\hline
   \multicolumn{7}{l}{\footnotesize$^a:$ $E_2$ contrib. from I--VI, represented with IC}
\\ \multicolumn{7}{l}{\footnotesize$^b:$ $E_2$ contrib. from VII--VIII, represented with MPSPT}
\end{tabular}
\end{table}

Fully uncontracted MRLCC3 is often very accurate\cite{Sharma2015mrlcc,Sharma2016} for a wide range of applications but is not practical for problems with large number of virtual orbitals. Compared to the uncontracted MRLCC3 energies, we expect three sources of error in our current implementation, the first is because we are only calculating second order energies, the second is due to the internal contraction and third is due to the fact that the reference wavefunction calculated using DMRG has inherent errors because low importance renormalized states are discarded. Using data in Table~\ref{tab:iclcc} we examine the error due to the first two reasons. By comparing the uncontracted MRLCC3 and uncontracted MRLCC2 energies it is clear that the third order energies are systematically more accurate than the second order energies, with errors in the atomization energies (AE) of the second order energies sometimes as large as 9 mE$_h$. The additional error introduced due to the use of internal contraction is minimal and the largest difference between the uncontracted and internally contracted MRLCC2 energies are only 1 mE$_h$. From the data it is evident that calculating the third order correction to the energy can significantly improve the quality of MRLCC results, but is outside the scope of the present work.

We have also calculated the ground state energies of the Chromium dimer using the fully uncontracted MRLCC2 with cc-pVTZ-DK basis set\cite{Balabanov2005} at a bond length of 1.68 \AA, relativistic effects are included using the second order Douglass-Kroll one electron Hamiltonian\cite{Reiher2004} and the reference wavefunction is obtained by performing a CASSCF (12e,12o) calculation\cite{Werner1985,Knowles1985}. Table~\ref{tab:lcc} shows the errors due to the use of internal contraction and an approximate reference wavefunction on the calculated MRLCC2 energies. By comparing the first two columns of the table, it can be seen that internal contraction causes the second order energy to be higher by 2.0 mE$_h$. The errors in energy differences are usually much smaller as demonstrated by the data in Table~\ref{tab:iclcc}. When the last three columns are compared one notices that by using approximate reference wavefunction parametrized using an MPS of M=50 an error of 3.8 mE$_h$ in the zeroth order energy $E_0$ is incurred. Here we have intentionally made the errors large by using an artificially small M, nevertheless such errors are not entirely uncommon when very large active spaces are treated using DMRG. The essential point is that the error in the second order energy due to a wrong reference is only 1.2 mE$_h$. In fact by using an even more approximate zeroth order state using an M=30 the error in the zeroth order energy is 30.7 mE$_h$ but the error in the second order energy is only 8.8 mE$_h$, with 5.1 mE$_h$ coming from the classes treated using internal contraction and the remaining 3.7 mE$_h$ from MPSPT calculations.


\subsection*{Computational Efficiency}
\begin{table*}
\caption{Walltime (in units of 1000 sec) for computing the various perturber classes' contributions to $|\Psi_1\rangle$. For classes treated with MPSPT, we show the time necessary to converge the energy to $\leq$1 mE$_h$.}\label{tab:timings}
\footnotesize
\begin{tabular}{lcccccccccccccccccc}
\hline\hline
& &  & && \multicolumn{3}{c}{Orbitals}&&\multicolumn{6}{c}{Internally contracted spaces}&~&\multicolumn{2}{c}{MPSPT}\\\cline{6-8}\cline{10-15}\cline{17-18}
Molecule~~~&   ~Sym~               & ~~State~~         &~~Active Space~~&  ~Basis~          &~~ $n_c$ & $n_a$ & $n_v$~~&Method&~~ I~~ & ~~II~~ & ~~III~~ & ~~IV~~ & ~~V~~ & ~~VI~~ &~& ~~VII~~ & ~~VIII\\
\hline
Cr$_2$  &   $D_{2h}$ &  $^1$A$_{1g}$ &(12e,12o)& V5Z & 18 & 12 & 276 & \tiny{MRLCC2}&  1.1 & 0.9 & 0.1 & 0.8 & 0.0 & 0.3 && 1.2 & 0.0 \\
Cr$_2$  &   $D_{2h}$ &  $^1$A$_{1g}$ & (12e, 30o)&V5Z & 18 & 30 & 258 & \tiny{MRLCC2}&  1.0 & 1.9 & 0.3 & 13.5 & 0.9 & 0.3 && 17.7 & 2.1 \\
Pentacene & $D_{2h}$ & $^1$A$_{1g}$  & (22e, 22o)& VDZ & 62 & 22 & 294 & \tiny{MRLCC2} & 16.8 & 5.0 & 1.2 & 2.5 & 1.2 & 6.2 && 10.9 & - \\
oxo-Mn(Salen) & $C_1$ & $^1$A & (28e, 22o) & VTZ & 55& 22 & 555 & \tiny{MRLCC2} & 42.6 & 33.6 & 0.8 & 29.7 & 0.2 & 1.8 &&  123.7& 33.75 \\

\\
Cr$_2$  &   $D_{2h}$ &  $^1$A$_{1g}$ &(12e,12o)& V5Z & 18 & 12 & 276 & \tiny{NEVPT2}&  0.0 & 0.0 & 0.0 & 0.0 & 0.0 & 0.0 && 1.2 & 0.0 \\
Cr$_2$  &   $D_{2h}$ &  $^1$A$_{1g}$ & (12e, 30o)&V5Z & 18 & 30 & 258 & \tiny{NEVPT2}&  0.0 & 0.0 & 0.0 & 0.0 & 0.0 & 0.0 && 17.7 & 2.1 \\
Pentacene & $D_{2h}$ & $^1$A$_{1g}$  & (22e, 22o)& VDZ & 62 & 22 & 294 & \tiny{NEVPT2} & 0.0 & 0.0 & 0.0 & 0.0 & 0.0 & 0.0 && 10.6 & - \\
Pentacene & $D_{2h}$ & $^1$A$_{1g}$  & (22e, 22o)& VTZ & 62 & 22 & 722 & \tiny{NEVPT2} & 0.5 & 0.8 & 0.0 & 0.0 & 0.3 & 0.2 && 29.2$^*$ & - \\
oxo-Mn(Salen) & $C_1$ & $^1$A & (28e, 22o) & VTZ & 55& 22 & 555 & \tiny{NEVPT2} & 0.0 & 0.2 & 0.0 & 0.1 & 0.0 & 0.0 &&  120.9& 37.4 \\
oxo-Mn(Salen) & $C_1$ & $^1$A & (28e, 22o) & VQZ & 55& 22 & 1101 & \tiny{NEVPT2} & 0.3 & 1.2 & 0.0 & 1.2 & 0.0 & 0.2 &&  487.2$^*$& 34.8 \\
\hline\hline
\end{tabular}
\begin{flushleft}\small{$^*$The virtual orbitals were split into four groups and the contribution of each group was calculated independently.}\end{flushleft}
\end{table*}
We have performed calculations on a variety of molecules with varying size of active space and basis sets to compare the cost of MRLCC2 and NEVPT2 calculations for the different perturber classes; the results are summarized in Table~\ref{tab:timings}.  All timings data were obtained on a single node containing two Intel\textsuperscript{\textregistered} Xeon\textsuperscript{\textregistered} E5-2680 v2 processors of 2.80 GHz each and 128 gigabyte memory. The table shows that there is substantial difference between MRLCC2 and NEVPT2 for the perturber classes that are treated using internal contraction, mainly because of the external-exchange operations (e.g. $\langle ab|cd\rangle t^{rs}_{ac}$) in MRLCC2 which are unnecessary in NEVPT2. The cost of such calculations scales as the fourth power of the number of virtual orbitals and with large basis sets quickly becomes the dominant cost. Here the largest MRLCC2 calculation was performed on oxo-Mn(Salen) with a cc-pVTZ basis\cite{Woon1993,Balabanov2005,dunningbasis} set which contains 555 virtual orbitals, whereas for NEVPT2 calculations very large basis sets containing over 1000 orbitals could be treated fairly easily. 

The MPSPT part of the calculation dominates the overall cost of the NEVPT2 calculation, but is of similar cost (within a factor of 3) as internal contraction in the MRLCC2 calculations. Even though the cost of performing MPSPT is about the same in NEVPT2 calculations as in MRLCC2 calculations, the NEVPT2 calculations can be stretched much further and performed more efficiently by using the fact that the perturbative contribution from each external orbital can be calculated independently of one another. We can perform simultaneous independent calculations where each one only correlates a subset of virtual orbitals. In the calculations on Pentacene with cc-pVTZ basis set and oxo-Mn(Salen) with cc-pVQZ basis set, there were respectively 722 and 1101 virtual orbitals, four independent MPSPT calculations were performed to treat the perturber class VII. It should be noted that, because 4-RDMs are not necessary large active spaces containing 30 orbitals (Cr$_2$ with (12e,30o) active space) can be treated using MPSPT.

\subsection*{oxo-Mn(Salen)}
\begin{figure}
\begin{center}
\includegraphics[width=0.5\textwidth]{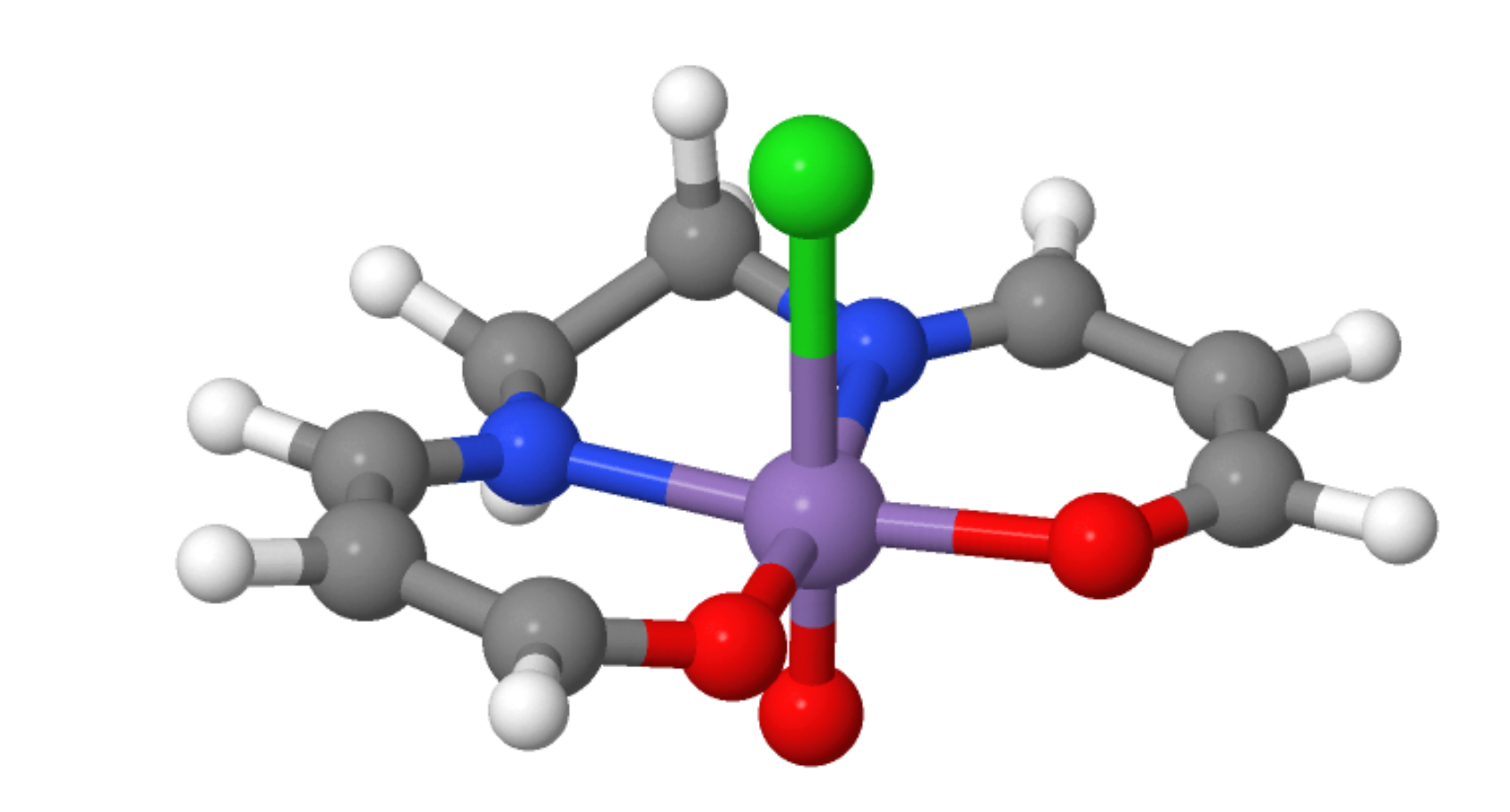}
\end{center}
\caption{The chlorine containing neutral oxo-Mn(salen) cluster used in this study. The geometry is taken from Ivanic et al.; it was optimized with (10e,10o)-CASSCF and a 6-31G* basis.}\label{fig:salen}
\end{figure}

Managanese salen derivatives like the Jacobson's catalyst\cite{Jacobsen1991,Zhang1990} are used for performing enetioselective epoxiation of olefins. The mechanism of the epoxidation is not clear, with various experiments providing evidence supporting at least three different possibilities\cite{Linker1997,Fu1991}. To provide further clarification, several theoretical studies\cite{C4CP00721B,Stein2016,Bogaerts2015,Ma2011,Sears2006,Ivanic2004,Wouters2014,Linde1999,Abashkin2001,Abashkin2004,Abashkin2001a} have tried to understand the electronic structure of these clusters. The chlorine containing neutral oxo-Mn(salen) model cluster studied in this work is shown in Figure~\ref{fig:salen}. Based on DFT and coupled cluster calculations\cite{Abashkin2001} it was shown that the cluster can be in a singlet, triplet and quintet state, with singlet being the most stable. The DFT calculations suggest that the reaction mechanism depend on the spin state of the cluster with singlet favoring the concerted mechanism which exclusively yields the cis isomer, whereas the triplet and the quintet favor the two-step mechanism which allows the formation of both the cis and trans isomers. Multireference \emph{ab initio} calculations\cite{Ivanic2004,Ma2011,Wouters2014,Sears2006,Stein2016} have also been performed by several groups to evaluate the relative energies of the three spin states. But these multireference calculations have failed to include dynamical correlation and thus their results need further confirmation.

Here we perform MRLCC and NEVPT2 calculations on the singlet and triplet ground states of the oxo-Mn(salen) using the optimized geometry obtained by Ivanic et al.\cite{Ivanic2004} Based on recommendation by Wouters et al.\cite{Wouters2014} a (28e,22o) active space was used to perform the DMRG-SCF calculations\cite{Zgid2008dmrgscf,Ghosh2008} where the DMRG calculation during each iteration was performed with an M of 2000. HOMO-13 to LUMO+7 canonical Hartree Fock orbitals were included in the active space in the first iteration.  StackBlock\cite{Sharma2015a,sharma:124121,Olivares-Amaya2015} was interfaced with the Pyscf program package\cite{Sun2015LibcintAE} to carry out these calculations. Remarkably, our DMRG-SCF calculations for the singlet state with the cc-pVDZ basis set converged to -2251.7991 E$_h$ which is about 50 mE$_h$ lower than the energy obtained by Wouters et al.\cite{Wouters2014} We are not certain why the difference between the two energies is so large, one possibility is that their active space calculation converged to a local minimum.  Our numbers are in good agreement with the recent results published by Stein and Reiher\cite{Stein2016}, where they have used a slightly smaller active space of (26e,21o) and obtained slightly higher energy of -2251.7963 E$_h$.

\begin{figure}
\begin{center}
\includegraphics[width=0.5\textwidth]{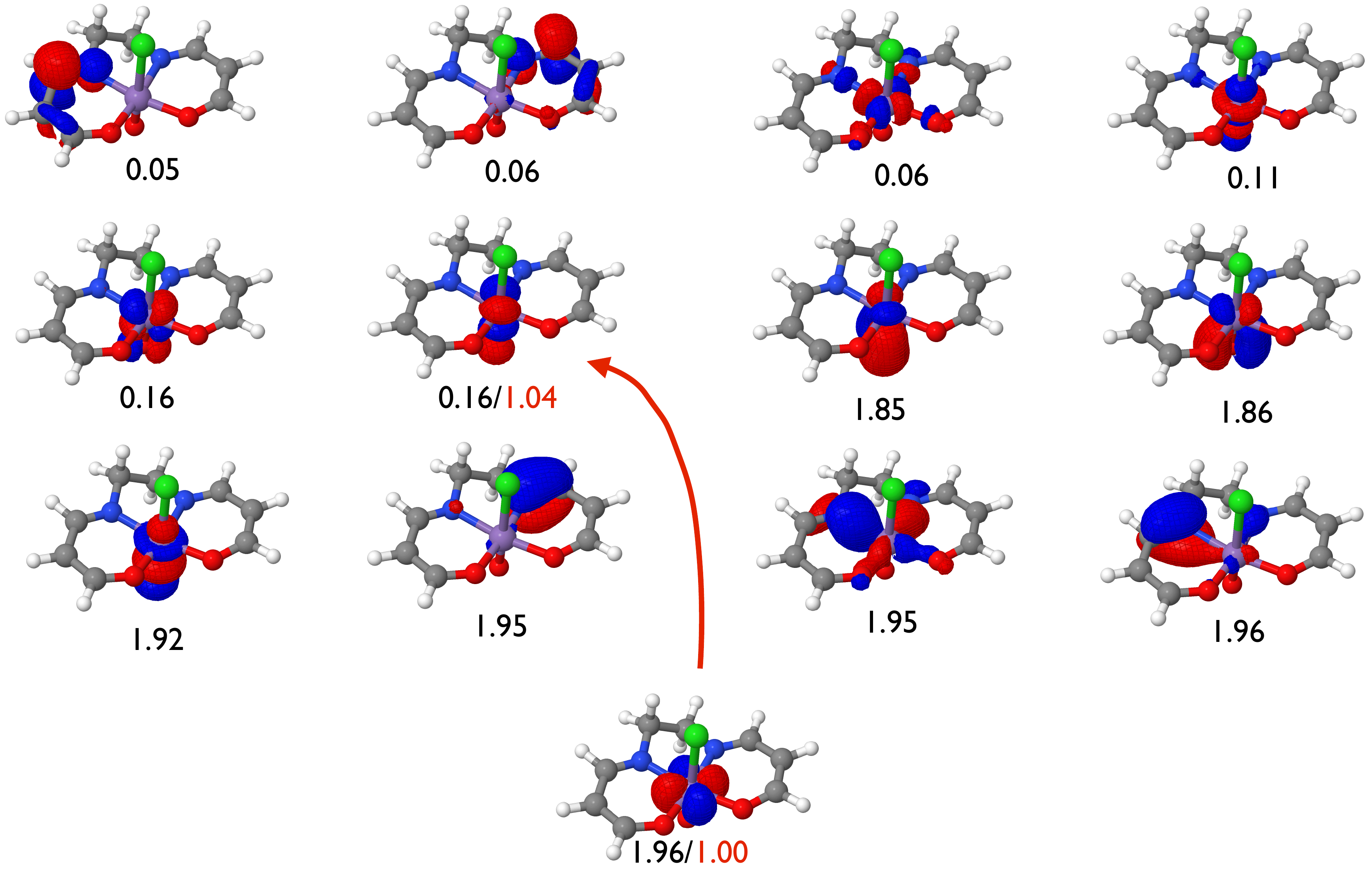}
\end{center}
\caption{The natural orbitals with the occupation numbers (numbers in black) greater than 0.02 or less than 1.98 of the singlet state are obtained using CASSCF (28e,22o) calculation with a cc-pvDZ basis set. In the triplet state an electron from the doubly occupied $d_{x^2-y^2}$ orbital gets excited to the empty $\pi^*$ orbital which is shown schematically with the red arrow. The occupation numbers of the $d_{x^2-y^2}$ orbital and $\pi^*$ orbitals change from 1.96 and 0.16 in the singlet state to 1.00 and 1.04 (shown in red) in the triplet state respectively.}\label{fig:natorbs}
\end{figure}

\begin{table*}
\caption{The energies (E+2251) in units of E$_h$ obtained using various methods and basis sets shown in the table. Comparing the NEVPT2/MRLCC results calculated at VDZ and VTZ basis set shows the use of the VDZ basis set results in large errors in the calculated singlet triplet splittings. The energy differences seem to have converged with basis set by VTZ basis set. The spin gap calculated using MRLCC and NEVPT2 calculations are in agreement with each other when VTZ basis set is used.}\label{tab:salen}
\begin{tabular}{lcccccccccccc}
\hline
&&\multicolumn{3}{c}{VDZ}&&\multicolumn{3}{c}{VTZ}&&\multicolumn{3}{c}{VQZ}\\
State&&~~E$_0$&NEV&LCC~~&&~~E$_0$&NEV&LCC~~&&~~E$_0$&NEV&LCC\\
\cline{1-1}\cline{3-5}\cline{7-9}\cline{11-13}
$^1$A&&-0.7991&-3.0109&-3.2830&&-0.9957&-3.8437&-4.1303&&-1.0449&-4.1441&-\\
$^3$A&&-0.8002&-2.9990&-3.2600&&-0.9926&-3.8463&-4.1310&&-1.0418&-4.1481&-\\
$\Delta$E$_{T-S}$&&-0.0011&0.0118&0.0230&&0.0031&-0.0026&-0.0008&&0.0031&-0.0039&-\\
\hline
\end{tabular}
\end{table*}
The results of our calculations are presented in Table~\ref{tab:salen}, where it can be seen that the singlet-triplet splitting not only depends on the dynamical correlation but also depends quite sensitively on the size of the basis set. Interestingly the DMRG-SCF calculations with VDZ basis set predicts the triplet as being more stable than the singlet by 1.1 mE$_h$, which is in contradiction to the results obtained by Stein and Reiher where they obtained singlet as being more stable than the triplet by 0.6 mE$_h$. We don't expect our DMRG-SCF calculations performed with M=2000 (or their calculations performed with M=1000) to be accurate enough to resolve such small differences. Further there is a strong likelihood that the ground state triplet is nearly degenerate with the first excited triplet state\cite{Sears2006} which can easily explain the difference in the excitation energies. The splitting calculated using DMRG-SCF seems to converge with basis set when going from VTZ to VQZ basis set. The natural orbitals of the singlet state calculated using the VDZ basis set are shown in Figure~\ref{fig:natorbs}. The shape and the occupancy of the natural orbitals do not change substantially when the basis set is changed, signifying that the qualitative nature of the singlet state is captured already at the VDZ basis set. Comparing the occupation number of the singlet and triplet natural orbitals shows that one of the electrons from the doubly occupied $d_{x^2-y^2}$ orbital in the singlet state gets excited to the $\pi^*$ orbital in the triplet state. This transition is in agreement with other multireference calculations. 

The dynamical correlation is quite sensitive to the basis set, with VDZ basis set tends to strongly stabilize the singlet state in sharp contrast to the results calculated using the VTZ and VQZ basis set. The agreement between the NEVPT2 results with the VTZ and VQZ basis set is quite good and NEVPT2 results at the VTZ basis set are in good agreement with the MRLCC2 results with the same basis set. (We were unable to perform MRLCC2 calculations with VQZ basis set because of the size of the external space.)

From our work it is clear that the difference between the energy of the singlet and triplet states is relatively small and is more or less within the error bars of the methods being used. An interesting question is that of the reaction mechanism on the different potential energy surfaces, which can be further explored by calculating the barrier heights of the transition states using MRLCC2 and NEVPT2 methods; this will be the topic of a future publication.

\section{Conclusion and Outlook}\label{sec:c}
 Here we have shown that Celani-Werner scheme of internal contraction can be used to perform NEVPT2 and MRLCC2 calculations for problems where the reference wavefunction is given as an MPS following a DMRG-CI or a DMRG-SCF calculation. This program can be used to perform NEVPT2 calculations for active spaces of 30 orbitals or more and with about 1000 virtual orbitals and MRLCC calculations with similar active spaces and about 500 virtual orbitals. This lends us the ability to calculate dynamical correlations on organometallic clusters where so far only static correlation have been calculated using DMRG\cite{Kurashige2013,sharma2014}. 
 
 Our results show that we can improve the accuracy of the MRLCC theory by calculating the third order energy corrections. The third order energies can be calculated by evaluating \[E_3 = \sum_{I>J}2\langle\Psi_1^J|V |\Psi_1^I\rangle \], where  $|\Psi_1^I\rangle$ is the $I^{th}$ perturber class of the first order state. There are eight different perturber classes and the resulting 28 different matrix elements need to be calculated to evaluate the third order energy. Out of these, all the ones where both the $I^{th}$ and $J^{th}$ perturber classes are parametrized using internal contraction can be easily calculated using a maximum of 3-RDMs. The matrix elements, where the $I^{th}$ perturber class in calculated using MPSPT and the $J^{th}$ perturber class is calculated using internal contraction, cannot be calculated in a straightforward way and will be the topic of a subsequent paper.
 
 The size of problems treatable with this method can be significantly extended by using explicit correlation (F12 methods) and local correlations using pair natural orbitals (PNO). It has already been demonstrated that by performing DMRG calculations on an effective Hamiltonian obtained by canonical transcorrelation of Yanai\cite{yanai:084107}, spectroscopic accuracy can be obtained for the Be$_2$ dimer\cite{be2sandeep}. The canonical transcorrelation method\cite{yanai:084107} although universal (much like Torheyden's post F12 correction scheme\cite{torheyden:171103}), has a severe memory bottleneck and can only be used for relatively small basis sets\cite{Kersten2016}. A more efficient route for using F12-geminals in our method will be the approach used by Ten-no\cite{Ten-no2007} and later by Shiozaki et al.\cite{Shiozaki2010,shiozaki:034113} for multireference problems. PNO were used in the multireference problems for the first time by Fink et al.\cite{Fink1993}, and after a long lull PNO based methods have seen a huge revival due to work of Neese et al.\cite{Riplinger2013,Guo2016a}. Our next project will aim to make use of these recent developments by extending this method by combining it with F12-methods, PNOs. 
\begin{acknowledgements}
The calculations made use of the facilities of the Max Planck Society's Rechenzentrum Garching. S.S. acknowledges the supported from University of Colorado. 
\end{acknowledgements}

\end{document}